\RequirePackage{fixltx2e}

\documentclass[twocolumn,showpacs,preprintnumbers,amssymb]{revtex4}

\usepackage{graphics}
\usepackage{epsfig} 
\usepackage{hyperref}
\usepackage{graphics}
\usepackage{color}      
\usepackage{graphicx}
\usepackage{graphicx}
\usepackage{dcolumn}

\begin{document}

\title {Irreversible Brownian heat engine}
\author{Mesfin Asfaw  Taye}
\affiliation {Department of Physics, California State University\\ Dominguez Hills, California, USA }

\begin{abstract}

We model a Brownian heat engine as a Brownian particle that hops in a periodic ratchet potential where the ratchet potential is coupled with  a  linearly decreasing background temperature. It is shown that  the efficiency of such Brownian heat engine is far from Carnot efficiency even at  quaistatic limit.  At quasistatic limit, the efficiency  of the  heat engine approaches  the efficiency of endoreversible engine  $\eta=1-\sqrt{{T_{c}/T_{h}}}$  \cite{c18}. On the other hand,  the maximum power efficiency  of the engine approaches $\eta^{MAX}=1-({T_{c}/T_{h}})^{1\over 4}$. Moreover, the dependence of the  current as well as  the efficiency   on the model parameters is    explored  analytically by omitting the heat exchange via the kinetic energy.  In this case  we show that the optimized efficiency always lies between the efficiently at quaistatic limit  and the efficiency at maximum power. On the other hand, the efficiency  at maximum power  is always less than  the optimized efficiency since the fast motion of the particle comes at the expense of the energy cost.
If one includes the heat   exchange  at the boundary of the heat baths, the  efficiency of the engine  becomes much smaller than the Carnot efficiency. In addition, the dependence for  the  coefficient of performance of the refrigerator on the model parameters is explored by including the heat exchange  via the potential and kinetic energy.
We show  that such  a Brownian heat engine has a higher  performance
when acting as a refrigerator  than when operating as a device subjected to  a piecewise constant temperature. The role of time on the performance of the motor is also explored via numerical simulations.  Our numerical results depict that  the time $t$ as well as  the external load dictate  the direction of the particle velocity. Moreover  the performance of the heat engine improves with time. At large $t$ (steady state),the velocity, the efficiency and the  coefficient of performance of the refrigerator attain their maximum value.

\end{abstract}
         
\maketitle
\section{Introduction} 

The study of  noise-induced transport feature of micron and nanometer sized particles  is  vital 
for a better understanding of the nonequilibrium statistical physics \cite{c8,cc8}.  These micron and nanometer sized particles attain  a unidirectional motion when they are exposed to a  potential where the potential itself is subjected to 
 spatial or temporal symmetry breaking fields such as nonhomogeneous
temperature  \cite{c1,c2,cc10,cc11,cc12,c3,c4,c5,c6,c7}.  Earlier,  considering  a Brownian particle arranged to move along a flashing or rocking ratchet, the  dependence of the  unidirectional current  on model parameters is studied  by  P. Reimann, R. Bartussek, R. H\"aussler, and P. H\"anggi \cite{am7}.  On the other hand,  
several studies have been also conducted to understand the factors that affect the performance of
the Brownian engine that is driven by a spatially varying temperature 
 \cite{c9,c10,c11,c12,c13,c14,c15,c16,c17,c18,c19}.  More recently,    the effect of temperature on the performance
of the heat engine as well as on its mobility was  studied  by  us considering a viscous
friction that has an exponential temperature dependence as   proposed by Reynolds \cite{c20}. It is shown that 
for isothermal case the particle undergoes 
a unidirectional motion as long as a non-zero load is exerted. As one increases the thermal energy of the medium,
the particle mobility   steps up considerably. For nonhomogeneous temperature case, the direction of the velocity is 
 dictated by the load. We showed that the speed of the particle steps up when the temperature difference between the hot and cold reservoirs increases \cite{cc20}. 

Previous studies have also indicate  that when these microscopic devices operate at two thermal reservoirs $T_h$ and $T_c$, their efficiencies  and coefficient of performance of the refrigerator  approach Carnot efficiency  $ \eta_{CAR}=1-{T_c\over T_h}$ and Carnot refrigerator  $P_{ref}^{CAR}={T_c\over T_h-T_c}$  at quasistatic limit as long as the heat exchange via the kinetic energy is excluded. When the heat exchange via the kinetic  energy is included, Carnot efficiency and Carnot refrigerator are unattainable even at a quaistatic limit revealing that  Brownian heat engines are inherently irreversible.   The operation regime at quasistatic limit is the least desirable one since one should wait an infinite time for the engine to accomplish its task although the efficiency at this operation regime is the maximum one. Hence  the efficiency or the coefficient of performance of the refrigerator  evaluated at this regime serves as  upper bound and has  theoretical signiﬁcance. However, it is  irrelevant from a practical point of view since   real heat engines operate at finite time periods  and they are subjected to irreversibility as depicted in the works \cite{c9,c10,c11}. On other hand, previous study based on finite time thermodynamics  uncovered  that   for  endoreversible engine the efficiency  at maximum power reduces to $\eta=1-\sqrt{{T_{c}/T_{h}}}$.

Most of the previous  works  have focused on a Brownian heat engine that operates between two thermal reservoirs $T_h$ and $T_c$. It is also crucial to study the  role of thermal inhomogeneity on the performance of a Brownian heat engine.  In order to fill that gap,  recently we   modeled  a Brownian heat engine as a Brownian particle that hops in a periodic ratchet potential where the ratchet potential is coupled with  a  linearly decreasing background temperature \cite{c17}. We explored the thermodynamic  properties  of such a heat engine not only at a  quasistatic limit but also when it operates at finite time.

In this work,  we extend (reconsider) the previous work \cite{c17} and uncover far more results.  We first  explore how the velocity, the efficiency and performance of the refrigerator behave  as a function of the model parameters  by excluding the heat exchange  via the   kinetic energy.    It is shown that  the efficiency of such Brownian heat engine is far from Carnot efficiency even at  quaistatic limit.  At quasistatic limit, the efficiency  of the  heat engine approaches  the efficiency of endoreversible engine  $\eta=1-\sqrt{{T_{c}/T_{h}}}$  \cite{c18}. On the other hand,  the maximum power efficiency  of the engine approaches $\eta^{MAX}=1-({T_{c}/T_{h}})^{1\over 4}$. Moreover  we show that the optimized efficiency always lies between the efficiently at quaistatic limit  and the efficiency at maximum power. On the other hand, the efficiency  at maximum power  is always less than  the optimized efficiency since the fast motion of the particle comes at the expense of the energy cost.
If one includes the heat   exchange at the boundary of the heat baths, the efficiency as well as the coefficient performance of the engine becomes much smaller than the Carnot efficiency or refrigerator. In addition, the dependence for the coefficient of performance of the refrigerator on the model parameters is explored.
We show that such a Brownian heat engine has a higher performance
when acting as a refrigerator than when operating as a device subjected to a piecewise constant temperature.

The role of time on the performance of the motor is also explored via numerical simulations.  Our numerical results depict that the velocity of the particle increases with time.  The external load as well as the rescaled time $t$  detects the direction of the particle velocity.  When $t$ is small, the net particle flow is towards the left direction. For large time, current reversal occurs and the particle flow towards the right direction.  The efficiency of the engine  explicitly  relies on time. As time increases, the efficiency increases. At steady state, it saturates to a constant value.   At small time $t$,  the efficiency is  much less than Carnot efficiency  showing that  the system  exhibits  irreversibility at small $t$.  The coefficient of performance of the refrigerator also  steps up  as time increases. As $t$ further  steps up, it converges to  a constant value.

The rest of the paper is organized as follows. In section II, we present the model and the derivation for the stead  state current. In section III, we explore the dependence for the velocity  on model parameters. The dependence   for  the efficiency and coefficient of performance of the refrigerator on the model parameters is  discussed in section IV. The short time behavior of the system is  discussed   in Section V.
Section VI deals with 
summary and conclusion.

\section{Model and steady state current}
\begin{figure} 
\epsfig{file=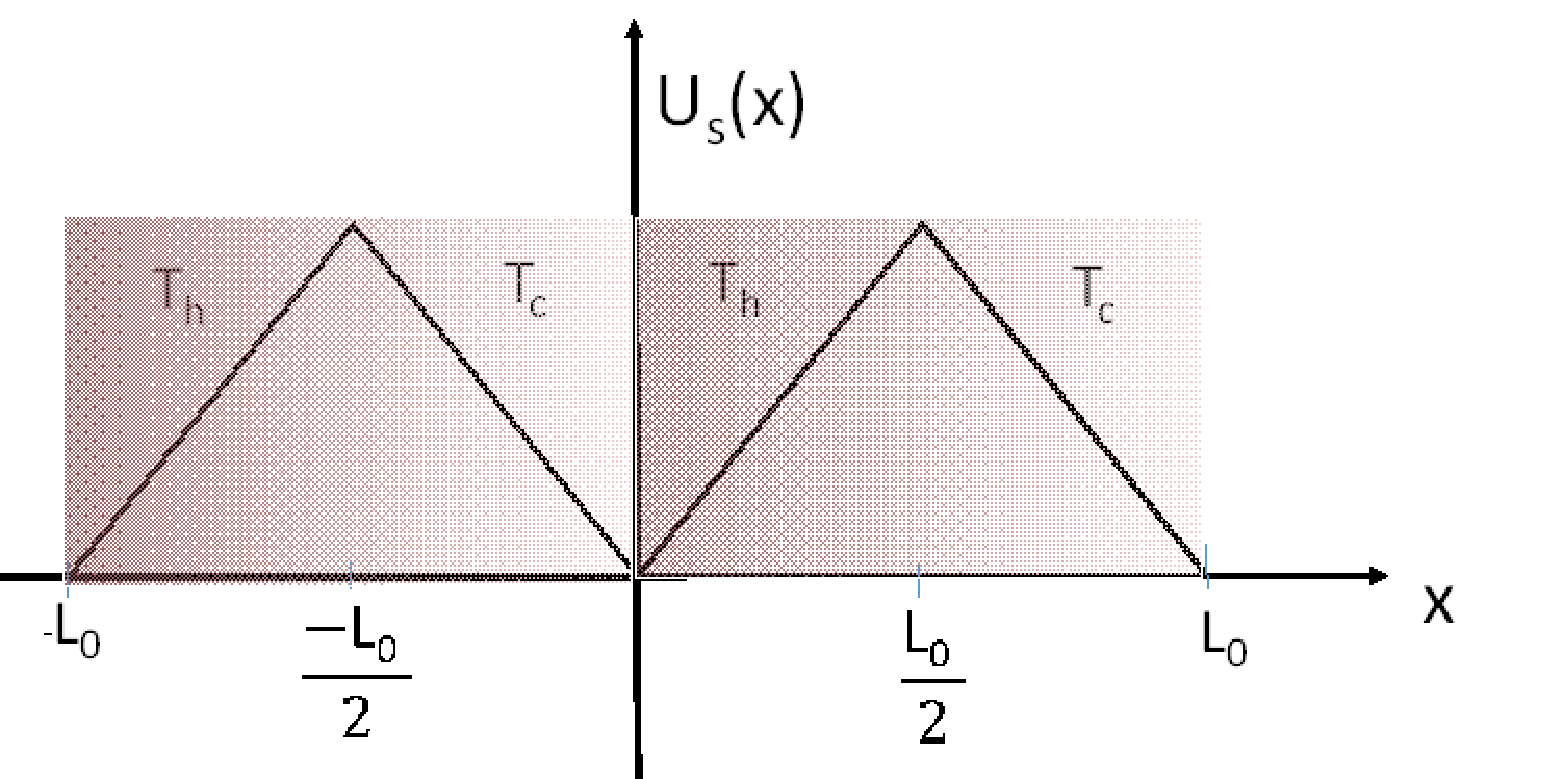,width=10cm}
\caption{ Schematic diagram for a Brownian particle in a piecewise linear  potential in the absence of external load. Due to  the thermal background kicks, the particle ultimately attains a steady state current (velocity) as long  a distinct temperature difference between the hot and the cold reservoirs is retained.}
\end{figure}

We consider  a  Brownian particle  that moves    along the potential $U(x)=U_{s}(x)+fx$  where $f$ and $U_{s}(x)$ denote the load  and  ratchet potential, respectively.  The ratchet potential $U_{s}(x)$ 
 \begin{equation} 
  U_{s}(x)=\left\{\begin{array}{cl}
   2U_{0}[{x\over L_0}],&if~ 0< x \le L_0/2;\\
   2U_{0}[{-x\over L_0}+1],&if~ L_0/2 < x \le L_0;\end{array}\right.
   \end{equation}
   is  coupled  with   a  heat bath  that  
  decreases  from $T_{h}$ at $x=0$  to $T_{c}$ at $x=L_0$  along the  reaction coordinate   in the manner
 \begin{equation} 
 T(x)=\left\{{x(T_{c}-T_{h})\over L_0}+T_{h}\right\}.
   \end{equation}
Here $U_{0}$ and $L_0$ denote  the barrier height and the width of the ratchet potential, respectively.  The ratchet  potential has 
 a potential maxima  at  $x=L_0/2$ and potential minima at $x=0$ and $x=L_0$. 
The potential profile repeats itself such that $U_{s}(x+L_0)=U_{s}(x)$. 

In the high friction limit, the dynamics of the particle is governed by the Langevin equation.
The general stochastic Langevin equation 
 which  derived in the  pioneering work of Petter H\"anggi \cite{am1,am2}  can be  written as
\begin{eqnarray}
\gamma(x){dx\over dt}&=&-{\partial U(x)\over \partial x} -{(1-\epsilon)\over \gamma(x)}{\partial\over   \partial x}(\gamma(x)T(x))+ \\ \nonumber
&&\sqrt{2k_{B}\gamma(x) T(x)}\xi(t)
\end{eqnarray}
following the approach stated in the work \cite{am3}. The  Ito  and  Stratonovich interpretations correspond to the case where $\epsilon=1$  and $\epsilon=1/2$, respectively while the case  $\epsilon=0$ is   known as  the  H\"anggi   a post-point or transform-form interpretation. Here after we adapt the Langevin equation 
\begin{eqnarray}
\gamma(x){dx\over dt}&=&-{\partial U(x)\over \partial x} + \sqrt{2k_{B}\gamma(x) T(x)}\xi(t).
\end{eqnarray}  $\gamma(x)$ is the viscous friction, and $k_B$ is the Boltzmann's constant. 
 The random force  $\xi(t)$ is considered to be Gaussian and white noise satisfying
\begin{equation}
\left\langle  \xi(t) \right\rangle =0,~~~\left\langle \xi(t)  \xi(t+\tau) \right\rangle=
\delta(\tau).
\end{equation}
 The corresponding  Fokker Planck equation is given by 
       \begin{eqnarray} 
       {\partial P(x,t)\over \partial t}&=&{\partial\over  \partial x} ({1\over \gamma }[U'(x)P(x,t)+{\partial \over \partial x}
       (k_{B}T(x)P(x,t)) ])\nonumber \\&= &-{\partial J(x,t)\over \partial x}
   \end{eqnarray}
       where $P(x,t)$ is   the probability density of finding the particle at position $x$ and 
         time $t$, $J(x,t)$ denotes  the current.  
         Hereafter, the Boltzmann constant $k_{B}$ and $\gamma$  are  taken
         to be unity.   

We are interested  in the long time behavior of the system.  In this limit, the expression for 
the  constant current, $J$, is given by
          \begin{equation}   {1\over \gamma }\left[-U'(x)P^{s}(x)+{d \over dx}(T(x)
         P^{s}(x))\right]=J. \end{equation}  It is important to note that 
 in the absence of symmetry breaking fields, no net flow of particles is obtained.  Only in the presence of externally acting load or inhomogeneous temperature distribution, a unidirectional motion of particle is attainable. Hereafter,  for sake of simplicity, we introduce dimensionless   rescaled temperature $\tau=T_{h}/T_{c}$,  rescaled barrier height ${\bar U_{0}}=U_{0}/T_{c}$  and rescaled length ${\bar x}=x/L_0$.  Hereafter for simplicity the bar will be neglected.

The general expression
for the steady state current $J$   in any periodic potential with or without load is
reported in the works \cite{c9,c11}.  Following the same approach, we find  
the steady state current J  as 
 \begin{equation}   
 J= {-F\over G_{1}G_{2}+HF}.
  \end{equation} 
	where the expressions
for F, G1, G2, and H are given as 

\begin{widetext}
\begin{eqnarray}
F&=& -1+e^{-\frac{2 U_{2} \text{ln}\left[\frac{2}{1+\tau}\right]}{1-\tau}+\frac{2 U_{1} \text{ln}\left[\frac{1+\tau}{2 \tau}\right]}{1-\tau}}, \\
G_1&=&\frac{1-4^{\frac{U_{1}}{1-\tau}} \left(\frac{\tau}{1+\tau}\right)^{\frac{2 U_{1}}{1-\tau}}}{2 U_{1}}+  \nonumber \\ & &
\frac{2^{-1+\frac{2
U_{1}}{1-\tau}} \left(\frac{1+\tau}{\tau}\right)^{-\frac{2 U_{1}}{1-\tau}} \left(-1+4^{\frac{U_{2}}{1-\tau}} \left(\frac{1}{1+\tau}\right)^{\frac{2 U_{2}}{1-\tau}}\right)}{U_{2}},
    \\ 
    G_2&=&\frac{1}{2} \left(\frac{2 \tau}{-1+\tau-2 U_{1}}-\frac{4^{\frac{U_{1}}{-1+\tau}} \left(1+\frac{1}{\tau}\right)^{-\frac{2 U_{1}}{-1+\tau}}
(1+\tau)}{-1+\tau-2 U_{1}}\right) +  \nonumber \\ & &
{1\over 2}\left(\frac{4^{\frac{U_{1}}{-1+\tau}} \left(1+\frac{1}{\tau}\right)^{-\frac{2 U_{1}}{-1+\tau}} \left(1+\tau-2^{1+\frac{2 U_{2}}{-1+\tau}}
\left(\frac{1}{1+\tau}\right)^{\frac{2 U_{2}}{-1+\tau}}\right)}{-1+\tau+2 U_{2}}\right),
\\
   H&=&T_{1}+T_{2}(T_{3}+T_{4}+T_{5}),
    \\
   T_1&=&\frac{\tau \left(-1+4^{\frac{U_{1}}{1-\tau}} \left(\frac{\tau}{1+\tau}\right)^{\frac{2 U_{1}}{1-\tau}}\right)+U_{1}}{2 U_{1}
(1-\tau+2 U_{1})},
\\
    T_{2}&=&2^{-2+\frac{2 (U_{1}+U_{2})}{1-\tau}} \left(\frac{1+\tau}{\tau}\right)^{-\frac{2 U_{1}}{1-\tau}},
     \\
    T_{3}&=&\frac{2^{\frac{1-\tau-2 (U_{1}+U_{2})}{1-\tau}} \left(\frac{1+\tau}{\tau}\right)^{\frac{2 U_{1}}{1-\tau}}}{1-\tau-2 U_{2}}+\frac{2
\tau \left(-4^{-\frac{U_{2}}{1-\tau}}+\left(\frac{1}{1+\tau}\right)^{\frac{2 U_{2}}{1-\tau}}\right)}{(-1+\tau-2 U_{1}) U_{2}},
 \\
T_{4}&=&\frac{2^{-\frac{2 U_{1}}{1-\tau}} (1+\tau) \left(\frac{1+\tau}{\tau}\right)^{\frac{2 U_{1}}{1-\tau}} \left(-2^{-\frac{2 U_{2}}{1-\tau}}+\left(\frac{1}{1+\tau}\right)^{\frac{2
U_{2}}{1-\tau}}\right)}{(1-\tau+2 U_{1}) U_{2}},
 \\
  T_{5}&=&-\frac{2^{-\frac{2 U_{1}}{1-\tau}} (1+\tau) \left(\frac{1+\tau}{\tau}\right)^{\frac{2 U_{1}}{1-\tau}} \left(-2^{-\frac{2 U_{2}}{1-\tau}}+\left(\frac{1}{1+\tau}\right)^{\frac{2
U_{2}}{1-\tau}}\right)}{(1-\tau-2 U_{2}) U_{2}}.
\end{eqnarray}

\end{widetext}

Here  $U_{1}=U_{0}+f/2$  and $U_{2}=U_{0}-f/2$.
	The expression for the velocity is then given by $V=LJ$.

	\section{The mobility of the particle}
	
	\begin{figure}[ht]
\centering
{
    \includegraphics[width=6cm]{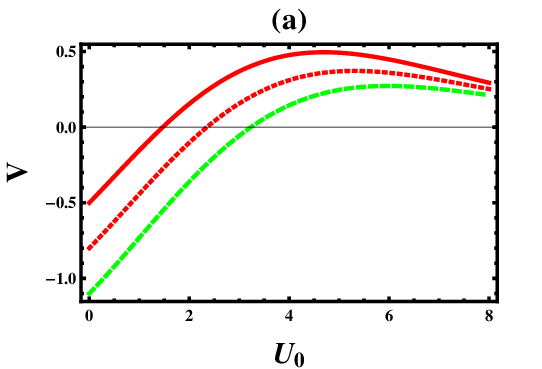}}
\hspace{1cm}
{
    \includegraphics[width=6cm]{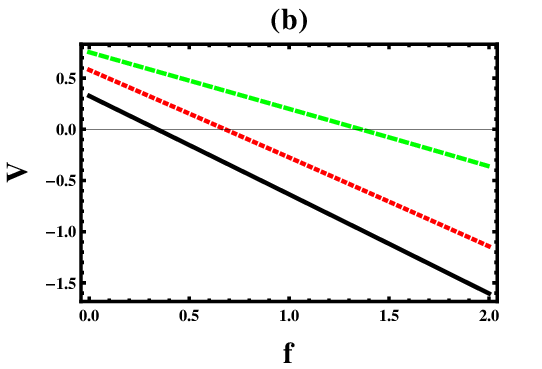}
}
\caption{ (Color online)(a) The velocity   $V$  as a function of $U_{0}$ for the  parameter value of $\tau=2.0$. The parameter $\lambda$ is fixed as $\lambda=0.5$, $\lambda=0.8$ and $\lambda=1.1$  from top to bottom, respectively.
 (b)  The velocity   $V$  as a function of  $\lambda$  for the  parameter value of $\tau=2.0$. The parameter $U_{0}$ fixed as $U_{0}=4.0$, $U_{0}=2.0$ and $U_{0}=1.0$  from top to bottom, respectively. } 
\label{fig:sub} 
\end{figure}
  
	\begin{figure}[ht]
\centering
{
    \includegraphics[width=6cm]{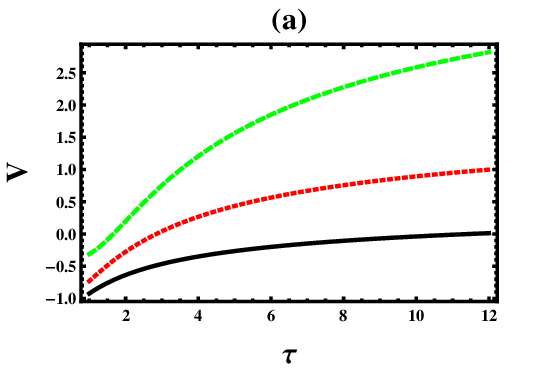}}
\hspace{1cm}
{
    \includegraphics[width=6cm]{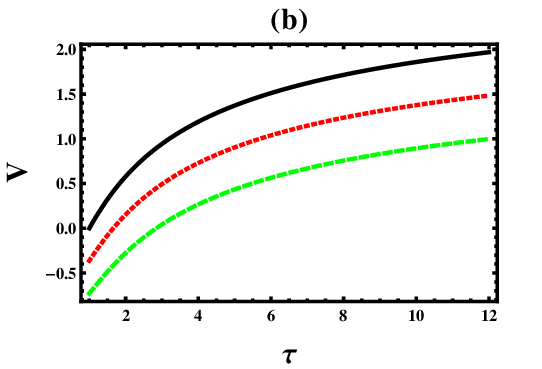}
}
\caption{ (Color online)(a) $V$  as a function of $\tau$ for the  parameter value of   $\lambda=1.0$. The parameter $U_{0}$ fixed as $U_{0}=4.0$, $U_{0}=2.0$ and $U_{0}=1.0$  from top to bottom, respectively.
 (b)  $V$  as a function of $\tau$ for the  parameter value of   $U_0=2.0$. The parameter $\lambda$ fixed as $\lambda=0.0$, $\lambda=0.5$ and $\lambda=1.0$  from top to bottom, respectively. } 
\label{fig:sub} 
\end{figure}

Various  studies on Brownian heat engine that operates  on the reaction coordinate that coupled with a  spatially varying temperature   have  depicted that the motor attains
a unidirectional motion as long as a distinct temperature difference is retained
along the potential.  The dependence  of  the steady state current or the velocity  on  the barrier  height $U_{0}$    can be  explored by 
exploiting Eq. (8).
 One can see that in the absence of the ratchet potential,  the average velocity of the particle is zero, i.e.; the velocity $V$  vanishes  when   $U_{0}\to 0$. In the limit  $U\to \infty$, $V\to 0$ which is expected because in the high barrier limit,  the particle encounters  a  difficulty of surmounting the high potential barrier.   In the presence of a load, the engine exhibits an
intriguing dynamics where  the magnitude of the load dictates  the  direction of
the particle flow.  The steady state current is zero at stall load 
 \begin{equation}   
f_{s} = {2U_{0}\over L}{\ln\left[{4\tau \over (1+\tau)^2}\right]\over \ln\left[{1\over \tau}\right]}.
  \end{equation}

The velocity   $V$  as a function of $U_{0}$ for the  parameter value of $\tau=2.0$ is plotted in Fig. 2a  for fixed  $\lambda=0.5$, $\lambda=0.8$ and $\lambda=1.1$  from top to bottom, respectively.  The figure depicts  that the velocity  steps up as $U_0$ increases   and at  a certain  potential height, the  velocity attains  its maximum value and it decreases  as the potential further increases. 
 The velocity   $V$  as a function of  $\lambda$  for the  parameter value of $\tau=2.0$, $U_{0}=4.0$, $U_{0}=2.0$ and $U_{0}=1.0$   is shown in Fig. 2b.  when $f< f_s$, $V>0$  in this regime the model acts as a heat engine  while $f> f_s$, $V<0$ in this case the model  function as a refrigerator.  
The magnitude of the steady state current also  strictly relies on  the rescaled   temperature $\tau$. When $\tau$ steps up,  the tendency of the particle in
the hotter bath to reach the top of the ratchet potential hill
increases than the particle in the colder reservoir. This leads
to an increase in the current $J$ or the drift velocity $V$ as
shown in Figs. 3a and 3b.  In Fig. 3a, we plot $V$  as a function of $\tau$ for the  parameter value of   $\lambda=1.0$. The parameter $U_{0}$ is  fixed as $U_{0}=4.0$, $U_{0}=2.0$ and $U_{0}=1.0$  from top to bottom, respectively.  The figure  depicts that the velocity increases as $\tau$ and  $U_{0}$ increase. The dependence for the velocity  on 
  $\tau$  is also explored in Fig. 3b  for the  parameter value of   $U_0=2.0$. The parameter $\lambda$ is fixed as $\lambda=0.0$, $\lambda=0.5$ and $\lambda=1.0$  from top to bottom, respectively. The figure exhibits that the velocity increases with $\tau$ and it decreases as the load increases.

	\section{The performance  of the motor}

\subsection{Energetics of the motor} 

The expressions for the work done by the Brownian particle as well as the amount heat taken from the hot bath and the  amount of heat given to the cold reservoir  can be derived in terms of the stochastic energetics discussed in the works \cite{am4,am5,am6}.  Let us first omit the heat dissipation via friction.   The heat taken from any heat bath can be evaluated  via \cite{am4,am5}   
${\dot Q}  =\left\langle \left(-\gamma(x){\dot x}+ \sqrt{2k_{B}\gamma(x) T(x)}\right).{\dot x}\right\rangle
  $
while the work done by the Brownian particle against the load is given by 
${\dot W}  =\left\langle f{\dot x}\right\rangle.
$
We can also find the expression for the input heat  $Q_{in}^s$  and $W^s$ \cite{am6} as 
\begin{eqnarray}
Q_{in}^s & =&\int_{0}^{L_{0}/2}\left(-\gamma(x){\dot x}+ \sqrt{2k_{B}\gamma(x) T(x)}\right)dx \\ \nonumber
&=&\int_{0}^{L_{0}/2}\left[\left({2U_{0}\over L_{0}}\right)+f\right]dx \\ \nonumber
&=&U_{0}+{fL_{0}\over 2}.
  \end{eqnarray}
Here the integral is evaluated in the interval of $(0,L_{0}/2)$ since the particle has to get a minimal amount of heat input from the heat bath located in the left side of the ratchet potential to surmount  the potential barrier.  The work done is also given by
\begin{eqnarray}
W^s &=&\int_{0}^{L_{0}}fdx=fL_{0}.
  \end{eqnarray}
	The first law of thermodynamics states that $Q_{in}^s-Q_{out}^s=W^s$  where $Q_{out}^s$  is the heat given to the colder heat bath.  Thus 
	$
Q_{out}^s =Q_{in}^s-W^s=U_{0}-{fL_{0}\over 2}.
 $

If one includes the  heat dissipation via the  viscous friction,  the minimum heat dissipation occurs  when the motor  hops with a constant velocity ${\dot x}=V$ \cite{c14}. In this case, an extra amount of heat $Q_{in}^*$  has to be taken  from the hotter bath to overcome the  friction. 
The average work done against the viscous friction is given by 
\begin{eqnarray}
W^* &=&\int_{0}^{L_{0}}\gamma(x){\dot x}dx \\ \nonumber
&=&\gamma({VL_{0}})=\gamma({JL_{0}^2})\\ \nonumber
&=&F_{av}L_{0}
  \end{eqnarray}
where 
 the average force $F_{av}=\gamma({JL_{0}})$.
On the other hand on average the heat taken from the hotter bath to overcome the frictional force is given as 
\begin{eqnarray}
Q_{in}^* &=&\int_{0}^{L_{0}/2}F_{av} dx \\ \nonumber
&=&\gamma({VL_{0}\over 2})=\gamma({JL_{0}^2\over 2}).
  \end{eqnarray}
Since $Q_{in}^*-Q_{out}^*=W^*$  and  one finds
	\begin{eqnarray}
Q_{out}^* &=&Q_{in}^*-W^*=-\gamma({JL_{0}^2\over 2}).
  \end{eqnarray}
In addition, ${1\over 2} { k_{B}(T_{h}-T_{c})}$   amount of heat per cycle is
transferred from the hotter to the colder heat baths via the kinetic energy in one cycle. Thus for single Brownian particle crossing over the potential barrier,
the amount of heat energy taken from the hot reservoir in  one cycle  is given by $
Q_{in}=Q_{in}^s+ Q_{in}^*+{1\over 2}k_{B}(T_{h}-T_{c})
=\left(U_{0}+\gamma J{L_{0}^2\over 2}+f{L_{0}\over 2}+{1\over 2}k_{B}(T_{h}-T_{c})\right).
$
This does make sense  since  in one cycle, a minimum
$(U_{0}+\gamma J{L_{0}^2\over 2}+f{L_{0}\over 2})$ amount of heat  
is needed to overcome the viscous drag force $\gamma V/2$, the potential barrier
$U_{0}$  and the external load $f$. In addition, ${1\over 2} { k_{B}(T_{h}-T_{c})}$   amount of heat per cycle is
transferred from the hotter to the colder heat baths. The heat given to the cold reservoir  takes a form 
 $
Q_{out}=Q_{out}^s+ Q_{out}^*+{1\over 2}k_{B}(T_{h}-T_{c})
=\left(U_{0}-\gamma J{L_{0}^2\over 2}-f{L_{0}\over 2}+{1\over 2}k_{B}(T_{h}-T_{c})\right).
$ The work done   against the load and  the
viscous friction is given by 
$
 W=Q_{in}-Q_{out}=W^s+W^*=\gamma({JL_{0}^2})+fL_{0}.
$
If the motor acts as a refrigerator,  the net heat flow  to the cold heat bath is given by \cite{c10} $
Q_{c}=\left(U_{0}-\gamma J{L_{0}^2\over 2}-f{L_{0}\over 2}-{1\over 2}k_{B}(T_{h}-T_{c})\right).
$
Moreover the efficiency is given as
\begin{eqnarray}
\eta= W/Q_{in}.
\end{eqnarray}  The performance of the refrigerator is also given by 
\begin{eqnarray}
P_{ref}=Q_{out}/W^L
\end{eqnarray}
where $W^L=fL_{0}$.

\subsection{The  efficiency of the heat engine }
{\it The  heat exchange via the potential.\textemdash}  We now explore the dependence of the efficiency $\eta$   on the model parameters.
To start with we first look at how $\eta$   
depends on the barrier height and the rescaled temperature $\tau$  by omitting the heat exchange
via kinetic energy. The efficiency  $\eta$ as a function of $U_0$ is depicted in Fig. 4a for the  parameter values of  $\lambda=0.0$, $\tau=8.0$, $\tau=6.0$, $\tau=4.0$ and $\tau=2.0$  from top to bottom, respectively. The figure exhibits that $\eta$ decreases from its maximum (quasistatic ) value  as $U_{0}$  and  $\tau$ increases.    In the quasistatic limit $U_{0} \to 0$ ($J\to 0$), we find 
\begin{equation}   
\eta^* = 1-{ln\left[{1+\tau\over 2\tau}\right] \over ln\left[{2\over \tau+1}\right]}
  \end{equation}
  which is approximately equal to the 
 efficiency of the endorevesible heat engine $\eta_{CA}$ 
  \begin{equation}   
\eta_{CA}=1-\sqrt{1/\tau}
\end{equation}   
as long as  the temperature difference between the hot and the cold reservoirs is not large. 
  In order to appreciate this let us Taylor expand Eqs. (26) and (27)  around $\tau=1$ and after some algebra one gets 
  \begin{eqnarray}   
\eta^* &=& \eta_{CA}={\tau-1\over 2}-{3\over 8}(\tau-1)^2+\ldots \nonumber\\
&=&{\eta_{CAR}\over 2}+{\eta_{CAR}^2\over 8}+{\eta_{CAR}^3\over 96}+\ldots
 \end{eqnarray}
  which exhibits that both efficiencies are equivalent  in this regime. Here $\eta_{CAR}$ is the Carnot efficiency $\eta_{CAR}=1-\sqrt{1/\tau}$. As discussed in the work \cite{mu25}, it is still unknown why different model systems approach the Taylor expression shown above. Indeed $\eta_C$ and $\eta_{CA}$ still precisely agree even at higher temperature difference as shown in Fig. 6a.

	The efficiency  $\eta$ as a function of $U_0$ is plotted in Fig. 4a for the  parameter values of  $\lambda=0.0$, $\tau=8.0$, $\tau=6.0$, $\tau=4.0$ and $\tau=2.0$. The efficiency decreases as the barrier height increases.  When the magnitude of the rescaled temperature steps up, the efficiency of the motor monotonously increases. 
	The dependence of $\eta$ on the rescaled  temperature $\tau$ is depicted in Fig. 4b for the  parameter values of $\lambda=0.0$, $U_{0}=1.0$, $U_{0}=2.0$, $U_{0}=4.0$ and $U_{0}=8.0$. The figure exhibits that the efficiency steps up with $\tau$ and it decreases as the barrier height decreases.
\begin{figure}[ht]
\centering
{
    \includegraphics[width=6cm]{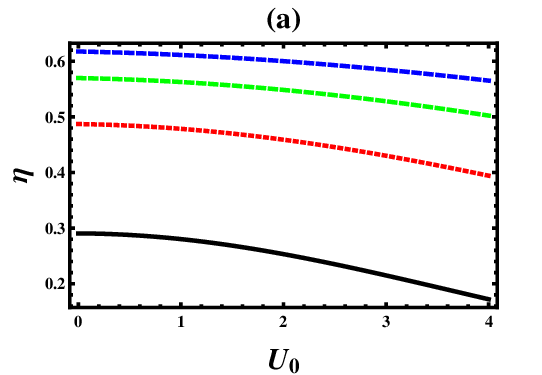}}
\hspace{1cm}
{
    \includegraphics[width=6cm]{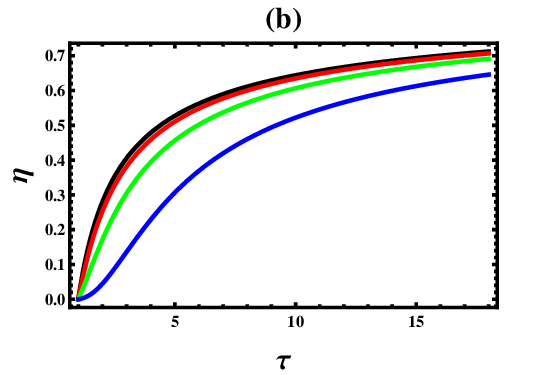}
}
\caption{ (Color online)(a) The efficiency  $\eta$ as a function of $U_0$ for the  parameter values of  $\lambda=0.0$, $\tau=8.0$, $\tau=6.0$, $\tau=4.0$ and $\tau=2.0$  from top to bottom, respectively.
 (b)  $\eta$  as a function of $\tau$ for the  parameter values of $\lambda=0.0$, $U_{0}=1.0$, $U_{0}=2.0$, $U_{0}=4.0$ and $U_{0}=8.0$  from top to bottom, respectively. } 
\label{fig:sub} 
\end{figure}

	In the presence of load,  the quasistatic limit of the engine corresponds to the case where the current approaches zero either from the heat engine side or from the refrigerator. The steady state current is zero at stall load  (see Eq. 18). 
  This  stall force  serves as
  a boundary that  demarcating the domain of operation of the engine. When $f<f_{s}$ the model acts as a heat engine while  as long as $f>f_{s}$  the model behaves  as refrigerator.  In quasistatic limit $J\to 0$, once again we find
$   
\eta_{CA}\approx  \eta_{C} = 1-{ln\left[{1+\tau\over 2\tau}\right] \over ln\left[{2\over \tau+1}\right]}$.
	\begin{figure}[ht]
\centering
{
    \includegraphics[width=6cm]{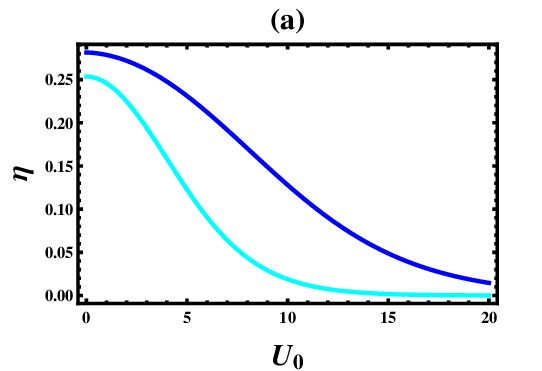}}
\hspace{1cm}
{
    \includegraphics[width=6cm]{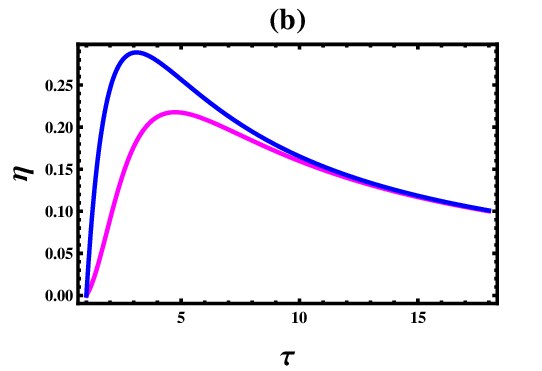}
}
\caption{ (Color online)(a) The efficiency  $\eta$ as a function of $U_0$ that plotted by considering the heat exchange via the kinetic energy. 
The  parameters are fixed as   $\lambda=0.0$, $\tau=4.0$ and  $\tau=2.0$  from top to bottom, respectively.
 (b)  $\eta$  as a function of $\tau$ for the  parameter values of $\lambda=0.0$, $U_{0}=2.0$ and  $U_{0}=6.0$ from top to bottom, respectively. } 
\label{fig:sub} 
\end{figure}

{\it The heat exchange via kinetic energy .\textemdash}  Let us now examine the thermodynamic property of the engine by
including the heat exchange via the kinetic  energy.
When the heat exchange via the kinetic energy is included Carnot efficiency
will not be obtained  even at the quasistatic limit. This is due
to the fact that the heat flow via kinetic energy is irreversible.

The dependence of  the efficiency on the model parameters is also examined by including the heat exchange via  kinetic energy. At 
quasistatic limit, the steady state efficiency  takes a form
\begin{equation}   
\eta = {\eta^* \over \Omega}
  \end{equation}
	where $\Omega$ is given by
	\begin{equation}   
 \Omega=\tau + {((-1 + \tau) Log[\tau])\over (Log[4] - 2 Log[1 + \tau])}
  \end{equation} 
	
Here  $0<{1\over \Omega}<1$, revealing that the efficiency can
never approaches the quasistatic efficiency $\eta^*$ that evaluated  by omitting the heat exchange via kinetic energy.

 The dependence of $\eta$ on the barrier height  is explored  in Fig. 5a. In the figure, the  parameters are fixed as   $\lambda=0.0$, $\tau=4.0$ and  $\tau=2.0$  from top to bottom, respectively. The figure depicts that  the efficiency  decreases    as 
the barrier height $U_ {0}$ decreases.  As shown in the same  figure, the efficiency decreases as $\tau$ decreases. In Fig. 5b, we plot $\eta$ as a function of $\tau$ for the  parameter values of $\lambda=0.0$, $U_{0}=2.0$ and  $U_{0}=6.0$ from top to bottom, respectively. The same figure depicts that  $\eta=0$ when  $\tau=1$ and it increases with $\tau$. The efficiency  attains an optimal value at a certain $\tau$ and it then decreases as $\tau$ further increases. $\eta$ also decreases as $U_0$ increases.

\subsection{Optimal and maximum power efficiency}

	The efficiency of the engine at maximum power $\eta^{MAX}$ is analyzed by substituting the values of $U_{0}$ and $\tau$ at which $J$ is maximum.  Fig. 6b (dotted line)  depicts  $\eta^{MAX}$ as a function of $\tau$  for fixed  $U_0=2$.  As it can be seen clearly $\eta^{MAX}$ is approximately  the same as the 
 efficiency  
  \begin{equation}   
\eta^{**}=1-(1/\tau)^(1/4)
\end{equation} 
at least in the small  $  \tau$ region. In fact $\eta^{MAX}$ and $\eta^{**}$ still precisely agree even at higher temperature difference as shown in Fig. 6b.

On the other hand, the optimized efficiency (OPT) is the efficiency where the competition between energy cost and fast transport is compromised \cite{c9}. Following the same approach as the work \cite{c9,cc20},  we optimize the function $\Omega=2W-{\tau-1 \over \tau}Q_{in}$.
In Fig. 7,  we plot the optimal efficiency as a function of $\tau$  (see the intermediate line). 
We want to stress that when the engine operates at quasistatic limit, the efficiency approaches $\eta^*$  (see the top line of  Figs. 7)  which is the maximum possible efficiency. At this operation regime, the particle  velocity is zero (zero energy cost). On the other hand, when the engine operates at maximum power, the velocity of the motor is maximum implies that the energy cost is higher and as a result, it becomes less efficient. However, the optimal efficiency (green dotted line) lies between $\eta^8$ and maximum power efficiency (red dotted line) as it can be seen in  Fig. 7.

The main message here is that   by selecting proper parameter space, we can control the operation  as well as  its task.  The operation regime at quasistatic limit is the least desirable one since one should wait an infinite time for the engine to accomplish its task although the efficiency at this operation regime is the maximum one; in other words, the system delivers a zero power. If one needs a motor that moves fast along the reaction coordinate, a proper value of $U_{0}$ that maximize the velocity can be selected as a possible model ingredient. A compromised effect  can be seen at optimal power efficiency regime.
\begin{figure}[ht]
\centering
{
    \includegraphics[width=6cm]{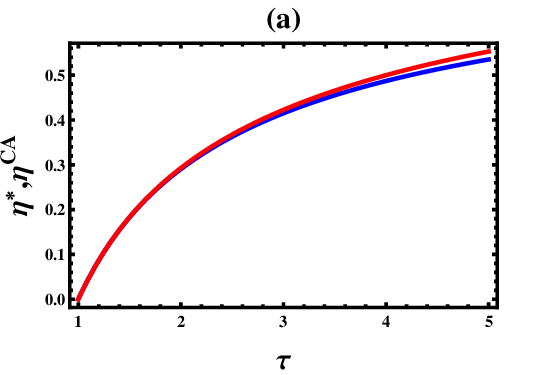}}
\hspace{1cm}
{
    \includegraphics[width=6cm]{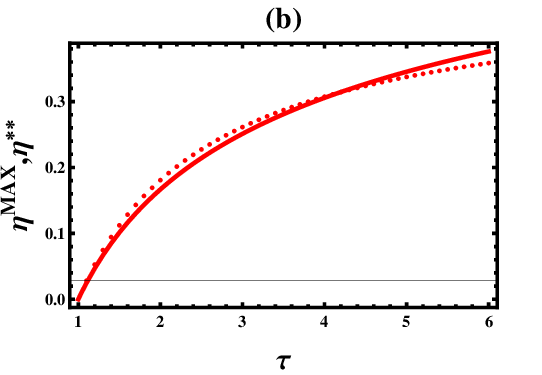}
}
\caption{ (Color online)(a) The quasistatic efficiency  $\eta^*$ (red line) and the efficiency $\eta^{CA}$ (blue line)  as a function of $\tau$.
 (b) The maximum efficiency  $\eta^{MAX}$ (dotted line) and $\eta^{**}$ (solid line) as a function of $\tau$. } 
\label{fig:sub} 
\end{figure}
\begin{figure} 
\epsfig{file=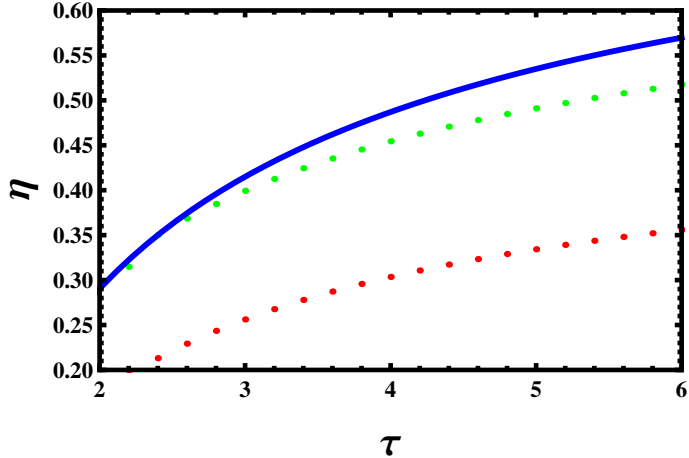,width=10cm}
\caption{ The efficiency $\eta$ as a function of $\tau$ for the parameter values of $\lambda = 0.5$ and $U_{0} = 8.0$. In the figure the top line stands for quasistatic efficiency $\eta^*$, the intermediate and the bottom line stand for optimum (OPT) and maximum power (MP) efficiencies, respectively. }
\end{figure}

\begin{figure}[ht]
\centering
{
    \includegraphics[width=6cm]{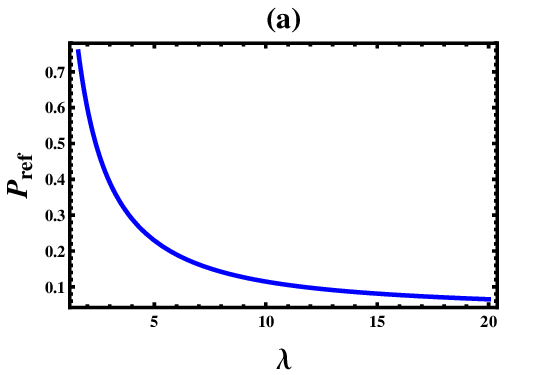}}
\hspace{1cm}
{
    \includegraphics[width=6cm]{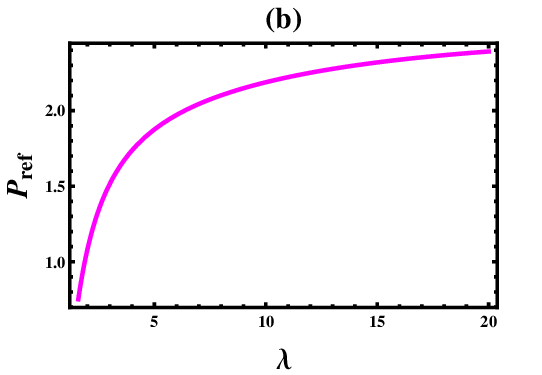}
}
\caption{ (Color online)(a) The  $P_{ref}$ as a function of $\lambda$.
 (b)  The  $P_{ref}$ as a function of $\lambda$  plotted  considering  the heat exchange via the kinetic energy.
 In both figures the parameters are fixed as   $U_0$ and  $\tau=6.0$} 
\label{fig:sub} 
\end{figure}

 \begin{figure}[ht]
\centering
{
    \includegraphics[width=6cm]{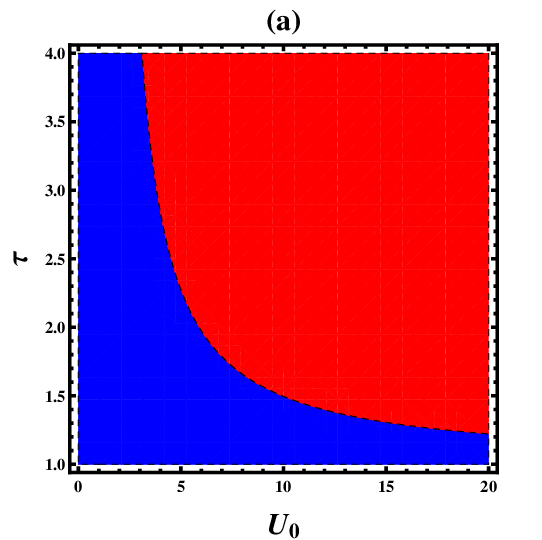}}
\hspace{1cm}
{
    \includegraphics[width=6cm]{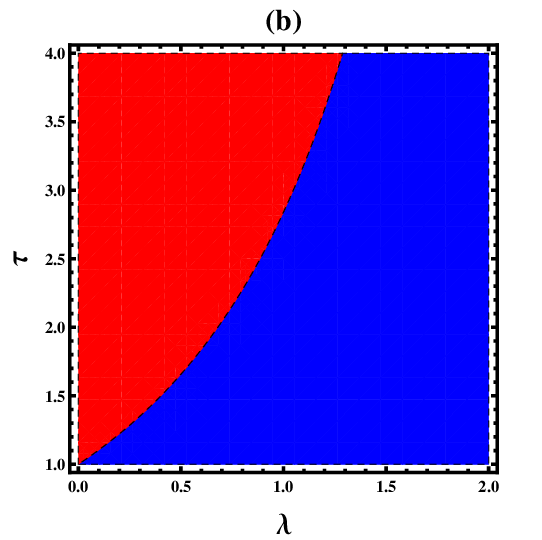}
}
\caption{ (Color online)(a) Phase diagram in   $ \tau$ and $U_{0}$ space   for a given $\lambda=2.0$.
 (b)  Phase diagram in   $ \tau$ and $\lambda$ space   for a given $U_0=2.0$.} 
\label{fig:sub} 
\end{figure}

\subsection{Coefficient of performance of the refrigerator}

The coefficient performance  of the refrigerator $P_{ref}$ is also explored as a function of the determinant  model parameters. At quasistatic limit,
$P_{ref}$ always approaches 
\begin{equation}   
P_{ref}^* = { \ln\left[{1\over 2}\left[{1\over \tau}+1\right]\right] \over \ln\left[{4\tau \over (1+\tau)^2}\right]}
  \end{equation}
which is much less than Carnot refrigerator.  As it can be readily seen that $P_{ref}^*$ decreases  as $\tau$ increases. 

The plot for 
$P_{ref}$ as a function of $\lambda$ is depicted in Fig. 8a for fixed    $U_0=2.0$ and  $\tau=6.0$. $P_{ref}$   decreases as the load  increases. The heat  exchange between the heat baths via the kinetic energy has also influence on $P_{ref}$. Figure 8b shows that the coefficient of performance  of the refrigerator increases as the load increases.  
Omitting the heat exchange via the kinetic energy,  a complete picture for the operation regions of the heat engine is obtained
by plotting the phase diagram in parameter space of $\tau$ and $U_{0}$ as 
shown in Fig. 9a. On the other hand, the phase diagram in parameter space of $\tau$ and $\lambda$
is plotted in Fig. 9b.
In both figures, the
region that marked red, the model works as a heat engine while in the region
that marked in blue the model acts as a refrigerator.

\section{Short time case}

In this work, we study the thermodynamic features of the engine  via numerical simulations. The numerical  results reveal the sensitivity of the performance of the thermal engine to the time $t$. The  operation regimes of the engine  are dictated by the operation time $t$. In the early particle relaxation period (small $t$), the engine neither acts as a heat engine nor as a refrigerator. This is because, when the system relaxation time is less than the time that the engine needs to perform work, the energy taken from the hot bath dissipates without doing any work. When $t$ further increases, depending on the parameter choice, the motor may work as a heat engine or as a refrigerator. Its performance is also an increasing function of $t$. Furthermore the engine depicts a higher efﬁciency or performance as a refrigerator at steady state regime. Moreover, we show that, when one omits the heat exchange via the kinetic energy,  Carnot refrigerator and Carnot efficiency are unattainable even when the system operates quasistatically at the steady state regime. The magnitude and the direction of the velocity are also controlled by $t$. 

\begin{figure}[ht]
\centering
{
    \includegraphics[width=6cm]{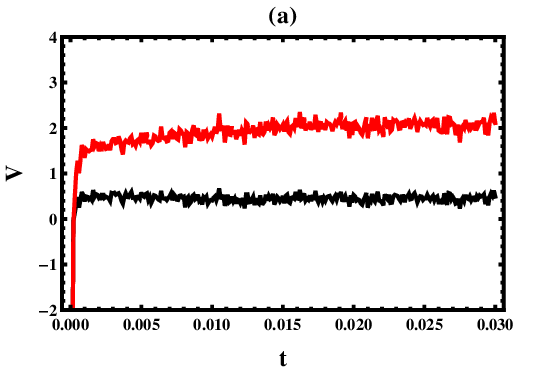}}
\hspace{1cm}
{
    \includegraphics[width=6cm]{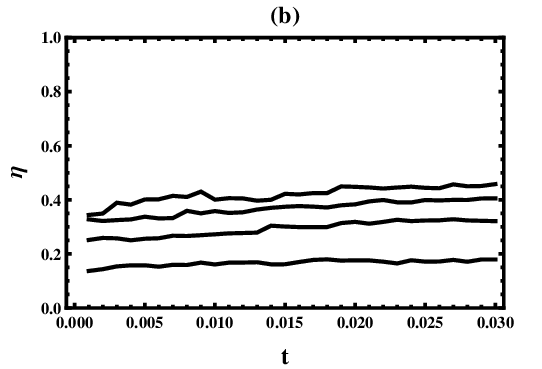}
}
\caption{ (Color online)(a) The velocity $V$ as a function of $t$ for parameter choice of $U_0=3.0$, $\lambda=0.2$. The rescaled temperature is fixed as  $\tau=8.0$ and $\tau=2.0$ from top to bottom.
 (b)   The efficiency  $\eta$ as a function of $t$ for parameter choice of $U_0=3.0$ and  $\lambda=0.2$. The rescaled temperature is fixed as  $\tau=8.0$, $\tau=6.0$, $\tau=4.0$ and $\tau=2.0$ from top to bottom. } 
\label{fig:sub} 
\end{figure}

The short time behavior of the velocity is quite sensitive to time $t$. In this case the magnitude and the direction of the velocity are dictated by $t$. This can be notably appreciated by looking at Fig. 10a. The figure depicts that, for very small $t$, the net particle  flow is from the colder to the hotter regions. As time increases,the magnitude of $V$ increases and stalls at certain time $t$. As time further steps up, the particle current gets reversed and the particle moves from the hotter to the colder reservoirs until its velocity saturates to a constant value.  The results obtained in this work also  agrees with our previous work \cite{c19}.  The  exact analytical work shown in the work  \cite{c19} also  uncovers  current reversal due to time $t$.  The efficiency $\eta$ as a function of time is also shown in Fig. 10b for parameter choice of  $U_0=3.0$ and  $\lambda=0.2$. The rescaled temperature is fixed as  $\tau=8.0$, $\tau=6.0$, $\tau=4.0$ and $\tau=2.0$ from top to bottom. One can see that  $\eta \ll \eta^*$  this is because the model is operating at finite time  and also   $\lambda=0.2 \ll f_s$.       The  same figure depicts that the efficiency steps up with time. As the rescaled temperature increases, the efficiency increases as expected.

Our numerical results depict that  for large time $t$ (approaching steady state), the  velocity (see Fig. 11)   as well as  the efficiency approach their steady state value. In Fig. 11, we plot $V$ as a function of $U_{0}$ for a given rescaled $t=100$, $\lambda=0.6$ and $\tau=2$. The numerically evaluated velocity (dotted line) coincides with the velocity evaluated via Eq. (18) (exact expression). Moreover, our analysis indicates that  
the coefficient of  performance of the heat refrigerator  improves with time. At large $t$ (steady state), the velocity, efficiency and coefficient of performance of the refrigerator attain their maximum value. 
 \begin{figure} 
\epsfig{file=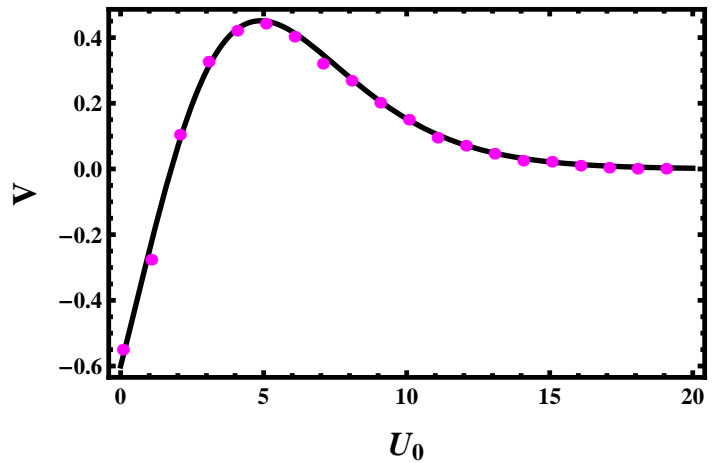,width=10cm}
\caption{ The efficiency $\eta$ as a function of $U_{0}$ for the parameter values of $\lambda = 0.6$, $t=100.0$ and $\tau = 2.0$. The numerically evaluated velocity (dotted line) coincides with the velocity evaluated via Eq. (18) (solid line).  }
\end{figure}

\section{Summary and conclusion } 
   
   In this work,  we consider   a Brownian  heat engine  that modeled as a particle hopping in a one-dimensional  periodic
ratchet  potential that  coupled with  a linearly decreasing  background temperature. Extending
  the previous work \cite{c17}, we  uncover far more results.  The dependence  of the velocity, the efficiency and performance of the refrigerator on the model parameters   is explored a by excluding the heat exchange  via the   kinetic energy.    We show that the efficiency of such Brownian heat engine is far from Carnot efficiency even at  quaistatic limit.  At quasistatic limit, the efficiency  of the  heat engine approaches  the efficiency of endoreversible engine  $\eta=1-\sqrt{{T_{c}/T_{h}}}$  \cite{c18}. On the other hand,  the maximum power efficiency  of the engine approaches $\eta^{MAX}=1-({T_{c}/T_{h}})^{1\over 4}$.  It is shown that 
	the optimized efficiency always lies between the efficiently at quaistatic limit  and the efficiency at maximum power. The efficiency  at maximum power  is always less than  the optimized efficiency since the fast motion of the particle comes at the expense of the energy cost. If one includes the heat   exchange at the boundary of the heat baths, the efficiency as well as the coefficient performance of the engine becomes much smaller than the Carnot efficiency or refrigerator. The dependence for the coefficient of performance of the refrigerator on the model parameters is  also explored.

Via  numerical simulations, we study 
the role of time on the performance of the motor.  The numerical results show  that the velocity of the particle increases with time.  The external load detects the direction of the particle velocity.  When the load is small, the net particle flow is towards the right direction and the model may act as a heat engine. For large load, current reversal occurs and the engine may work as a refrigerator.  The efficiency of the engine  explicitly  relies on time. As time increases, the efficiency increases. At steady state, it saturates to a constant value.   At small time $t$,  the efficiency is  much less than Carnot efficiency  showing that  the system  exhibits  irreversibility at small $t$.  The coefficient of performance of the refrigerator also  steps up  as time increases. As $t$ further  steps up, it converges to  a constant value.

 In conclusion,  the model of Brownian heat engine  which is presented in this work  serves as a guide in the construction of artificial microscopic heat engine  and also it is  crucial for fundamental understanding of the  nonequilibrium physics. We also believe that the present study   serves as a basic exemplar to study      the transport feature of biologically relevant systems such as polymers and membranes. 
  
   \section*{Acknowledgements }
I would like also to  thank Mulu Zebene  for her constant encouragement.


\begin{thebibliography}{60}
\bibitem{c8} P. H\"anggi, F. Marchesoni, and  F. Nori, Ann. Phys. (Leipzig) {\bf 14}, 51 (2005).
\bibitem{cc8} P. H\"anggi and F. Marchesoni, Rev. Mod. Phys. {\bf 81}, 387 (2009).
\bibitem{c1} T. Hondou and  K. Sekimoto, Phys. Rev. E {\bf 62}, 6021 (2000).
\bibitem{c2} A.G. Marin and  J.M. Sancho, Phys. Rev. E {\bf 74}, 062102 (2006).
\bibitem{cc10} N. Li, F. Zhan, P. H\"anggi, and B. Li, Phys. Rev. E {\bf 80}, 011125 (2009).
\bibitem{cc11} N. Li, P. H\"anggi, and B. Li, Europhysics Letters  {\bf 84}, 40009 (2008).
\bibitem{cc12} F. Zhan, N. Li, S. Kohler, and P. H\"anggi, Phys. Rev. E {\bf 80}, 061115 (2009).
\bibitem{c3} M. B\"{u}ttiker, Z. Phys. B {\bf 68}, 161 (1987).
\bibitem{c4} N.G. van Kampen, IBM J. Res. Dev. {\bf 32}, 107 (1988).
\bibitem{c5} R. Landauer, J. Stat. Phys. {\bf 53}, 233 (1988).
\bibitem{c6} R. Landauer, Phys. Rev. A {\bf 12}, 636 (1975).
\bibitem{c7} R. Landauer, Helv. Phys. Acta {\bf 56}, 847 (1983).
\bibitem{am7} P. Reimann, R. Bartussek, R. H\"aussler, and P. H\"anggi, Phys. Lett.  A {\bf 215}, 26 (1996).
\bibitem{c9}  M. Asfaw and M. Bekele, Eur. Phys. J. B {\bf 38}, 457 (2004).
\bibitem{c10} M. Asfaw and M. Bekele, Phys. Rev. E {\bf 72}, 056109 (2005).
\bibitem{c11} M. Asfaw and M. Bekele, Physica  A {\bf  384}, 346 (2007).
\bibitem{c12} M. Matsuo and S. Sasa, Physica A {\bf 276}, 188 (1999).
\bibitem{c13} I. Der\`enyi  and R.D. Astumian, Phys. Rev. E {\bf 59}, R6219 (1999).
\bibitem{c14} I. Der\`enyi, M. Bier, and R.D.  Astumian, Phys. Rev. Lett {\bf 83}, 903 (1999).
\bibitem{c15} J.M. Sancho, M. S. Miguel, and  D. D\"urr, J. Stat. Phys. {\bf 28}, 291 (1982).
\bibitem{c16} B.Q. Ai, H.Z. Xie, D.H. Wen, X.M. Liu, and  L.G. Liu, Eur. Phys. J. B {\bf 48}, 101 (2005).
\bibitem{c17} M. Asfaw, Eur. Phys. J. B {\bf 86}, 189 (2013).
\bibitem{c18} F. L. Curzon and B. Ahlborn, Am. J. Phys. {\bf 43}, 22 (1975).
\bibitem{c19} M. Asfaw, Phys. Rev. E {\bf 89}, 012143 (2014).
\bibitem{c20} O. Reynolds,  Phil Trans Royal Soc London { \bf 177}, 157 ((1886).
\bibitem{cc20} M. Asfaw and S. F.  Duki, Eur. Phys. J. B {\bf 88}, 322 (2015).
\bibitem{am1} P. H\"anggi,  Helv. Phys. Acta  {\bf 51}, 183 (1978).
\bibitem{am2} P. H\"anggi,  Helv. Phys. Acta  {\bf 53}, 491 (1980).
\bibitem{am3}J. M. Sancho, M. S. Miguel and D. Duerr,  J. Stat. Phys. {\bf 28}, 291  (1982).
\bibitem{am4} K. Sekimoto,  J. Phys. Soc. Jpn.  {\bf 66}, 1234 (1997).
\bibitem{am5} K. Sekimoto,  Prog. Theor. Phys. Suppl. {\bf 130}, 17 (1998).
\bibitem{am6}  M.  Matsuo, , Shin-ichi Sasa, Physica A {\bf 276}, 188  (2000).
\bibitem{mu25}  A. Calvo Hernandez, J. M. M. Roco, and A. Medina, Revista Mexicana de Fısica  {\bf 60}, 384  (2014).
\end{thebibliography}
\end{document}